\documentclass[preprint,12pt,authoryear]{elsarticle}

\usepackage{lineno,hyperref}
\modulolinenumbers[5]
\usepackage{algorithm}
\usepackage{amsmath}  
\usepackage{algorithmic}
\usepackage{amsfonts}
\usepackage{bm}
\usepackage{dsfont}
\usepackage{multirow}
\usepackage{threeparttable} 
\usepackage{booktabs,caption}
\usepackage{booktabs}
\usepackage{array}
\begin{document}
\begin{frontmatter}

\title{MGA-Net: A Novel Mask-Guided Attention Neural Network for Precision Neonatal Brain Imaging}


\author[inibica]{Bahram Jafrasteh\corref{mycorrespondingauthor}\fnref{fn1}}
\ead{jafrasteh.bahram@inibica.es}
\fntext[fn1]{Currently at Weill Cornell Medicine, Department of Radiology.}

\author[inibica,hpum]{Sim\'on Pedro Lubi\'an-L\'opez}
\ead{simonp.lubian.sspa@juntadeandalucia.es}
\author[inibica]{Emiliano Trimarco}
\author[inibica]{Macarena Rom\'an Ruiz}
\author[hpum]{Carmen Rodr\'iguez Barrios}
\author[inibica]{Yolanda Mar\'in Almagro}
\author[inibica,hpum,uca]{Isabel Benavente-Fern\'andez}
\ead{isabel.benavente@uca.es}

\cortext[mycorrespondingauthor]{Corresponding author}

\address[inibica]{Biomedical Research and Innovation Institute of C\'adiz (INiBICA) Research Unit, Puerta del Mar University, C\'adiz, Spain}
\address[hpum]{Division of Neonatology, Department of Pediatrics, Puerta del Mar University Hospital, C\'adiz, Spain}
\address[uca]{Area of Pediatrics, Department of Child and Mother Health and Radiology, Medical School, University of C\'adiz, C\'adiz, Spain}

\begin{abstract}
In this study, we introduce MGA-Net, a novel mask-guided attention neural network, which extends the U-net model for precision neonatal brain imaging. MGA-Net is designed to extract the brain from other structures and reconstruct high-quality brain images. The network employs a common encoder and two decoders: one for brain mask extraction and the other for brain region reconstruction. A key feature of MGA-Net is its high-level mask-guided attention module, which leverages features from the brain mask decoder to enhance image reconstruction. To enable the same encoder and decoder to process both MRI and ultrasound (US) images, MGA-Net integrates sinusoidal positional encoding. This encoding assigns distinct positional values to MRI and US images, allowing the model to effectively learn from both modalities. Consequently, features learned from a single modality can aid in learning a modality with less available data, such as US.
We extensively validated the proposed MGA-Net on diverse and independent datasets from varied clinical settings and neonatal age groups. The metrics used for assessment included the DICE similarity coefficient, recall, and accuracy for image segmentation; structural similarity for image reconstruction; and root mean squared error for total brain volume estimation from 3D ultrasound images. Our results demonstrate that MGA-Net significantly outperforms traditional methods, offering superior performance in brain extraction and segmentation while achieving high precision in image reconstruction and volumetric analysis. Thus, MGA-Net represents a robust and effective preprocessing tool for MRI and 3D ultrasound images, marking a significant advance in neuroimaging that enhances both research and clinical diagnostics in the neonatal period and beyond.
\end{abstract}


\end{frontmatter}

\section{Introduction}
Neuroimaging studies are crucial for advancing our understanding of brain development and functioning, particularly in the field of neuroscience. The preprocessing step directly influences the quality and effectiveness of subsequent analyses \citep{biessmann2011analysis}. Brain image preprocessing in neonates, and even more so in preterm-born neonates, represents a critical area of research given the vulnerable nature of their developing brains and the clinical implications of these studies \citep{hintz2015neuroimaging, hinojosa2017clinical}.
During the neonatal period, Magnetic Resonance Imaging (MRI) and 3D ultrasound (US) are the primary modalities for brain analysis. MRI provides high-resolution, detailed structural images of the brain, essential for detecting developmental anomalies and guiding therapeutic interventions \citep{counsell2019fetal}. In contrast, 3D US offers a safe, real-time, and accessible option for bedside monitoring of cerebral anatomy and vascular structures \citep{stanojevic2002three, beijst2020two}. These modalities complement one another, providing comprehensive assessments that neither can achieve alone. However, processing data from both sources poses significant challenges, necessitating sophisticated preprocessing techniques to manage diverse data types, enhance image quality, and prepare datasets for analysis. Developing such tools is crucial for advancing neonatal neuroimaging and ensuring the clarity, usability, and harmonization of imaging data.

Preprocessing Neuroimaging data is a multi-step process, with brain extraction, commonly known as skull stripping, plays an essential and foundational role. It involves the precise removal of non-brain tissues, such as skull, scalp and neck fat, from MRI scans, ensuring that only the brain structure remains for furthered analysis \citep{thakur2020brain}. 
This step is critical, not only for enhancing the quality of subsequent analysses but also for complying with privacy regulations like the Health Insurance Portability and Accountability Act of 1996 (HIPAA) and the General Data Protection Regulation of 2016 (GDPR). 
It significantly impacts a wide range of neuroimaiging diagnostics for both adults and neonates, including brain tumor and white matter lesion segmentation, cortical surface reconstruction, surgical interventions, neurodegeneration studies, radiation therapy planning, image registration, and predictions related to diseases like Alzheimer's disease and multiple sclerosis \citep{zhao2010automatic,gitler2017neurodegenerative,radue2015correlation}.

Manual brain delineation, the prevailing method for skull stripping, is a labor-intensive process and prone to variability, making it unsuitable for large-scale studies. This appraoch is time-consuming, susceptible to inconsistencies between different raters, and poses challenges to reproducibility. Due to these limitations, there has been a growing emphasis on developing automated approaches that mitigate the challenges of manual skull stripping. In particular, the emergence of deep learning techniques, notably Convolutional Neural Networks (CNNs), has revolutionized image segmentation tasks. Hwang et al. \citep{hwang20193d} demonstrated the potential of 3D CNNs in achieving accurate and efficient skull stripping in MRI datasets. Yu et al. \citep{yu2022generalizable} extended this work by introducing domain-adaptive CNNs, which can generalize across different imaging modalities and species. Hoopes et al. \citep{hoopes2022synthstrip} proposed SynthStrip, a learning-based tool that leverages synthetic data for skull stripping, offering greater accuracy across diverse MRI protocols. Meanwhile, weakly supervised models like those of Ranjbar et al. \citep{ranjbar2022weakly} have proposed for for handling small population. Pei et al. \citep{pei2022general} further demonstrated the use of ensemble neural networks for skull stripping, illustrating the growing capabilities of CNNs in achieving reliable and reproducible outcomes in brain extraction.

While numerous preprocessing methods for MRI images have been developed, most research has focused on images acquired using standardized protocols, often neglecting those obtained in less controlled conditions \citep{iglesias2023synthsr}. This oversight is particularly significant when it comes to MRI data from preterm neonates, especially those with pathological conditions, which remains largely unexplored \citep{iglesias2023synthsr}. 
Recent advancements in artificial intelligence have proposed methods to enhance MRI image quality, but these approaches do not specifically address the unique challenges associated with neonatal imaging. Similarly, in the realm of 3D ultrasound (US) imaging, the majority of research has concentrated on fetal ultrasound \citep{namburete2018fully}. There is a notable lack of studies dedicated to the preprocessing of neuroimaging data obtained from neonates and preterm-born neonates \citep{jafrasteh2023deep}. This gap in automated preprocessing methods for neonatal ultrasound hinders their application in broader neuroimaging studies \citep{moser2022bean}.
To address these shortcomings, there is a need for the development of advanced preprocessing tools tailored specifically to MRI and ultrasound data from neonates. By focusing on these areas, future research can improve the quality and utility of neuroimaging analyses for diverse neonatal populations.

Despite the potential of separate deep learning models to preprocess MRI and ultrasound (US) images, significant challenges arise primarily due to the scarcity of extensive datasets, especially for ultrasound imaging. Deep learning models require substantial amounts of data to ensure robust performance, and the limited availability of comprehensive US datasets severely restricts the effectiveness of models tailored to a single modality. Although training a unified deep learning model with paired MRI and US images might appear beneficial, this approach often encounters feasibility issues related to data limitations, making it impractical for widespread implementation. Given these challenges, there is a compelling need for a novel deep learning approach that effectively leverages the strengths of available data across different imaging modalities.

Recent advancements in mask-guided attention have shown enhanced performance across various tasks, such as image classification \citep{wang2021mask}, image re-identification \citep{cai2019multi}, and occluded pedestrian detection \citep{pang2019mask}. Despite these developments, there is limited research on applying mask-guided attention to image reconstruction, particularly for brain image preprocessing.
In this study, we introduce a novel mask-guided attention module designed specifically for preprocessing and reconstructing MRI and 3D neonatal ultrasound images. This module leverages brain masks to improve the quality of image reconstruction. Additionally, we incorporate sinusoidal positional encoding to enhance the model’s capability to process and integrate data from distinct modalities within a unified framework.

MGA-Net, our proposed model, performs essential preprocessing tasks, including brain extraction, bias field correction, and noise removal for MRI, as well as similar preprocessing steps and total brain volume estimation for 3D neonatal ultrasound images. This integrated approach addresses data scarcity issues and improves the model’s applicability and effectiveness in clinical settings.

\begin{figure}[htpb]
\centering
\includegraphics[width=1\linewidth]{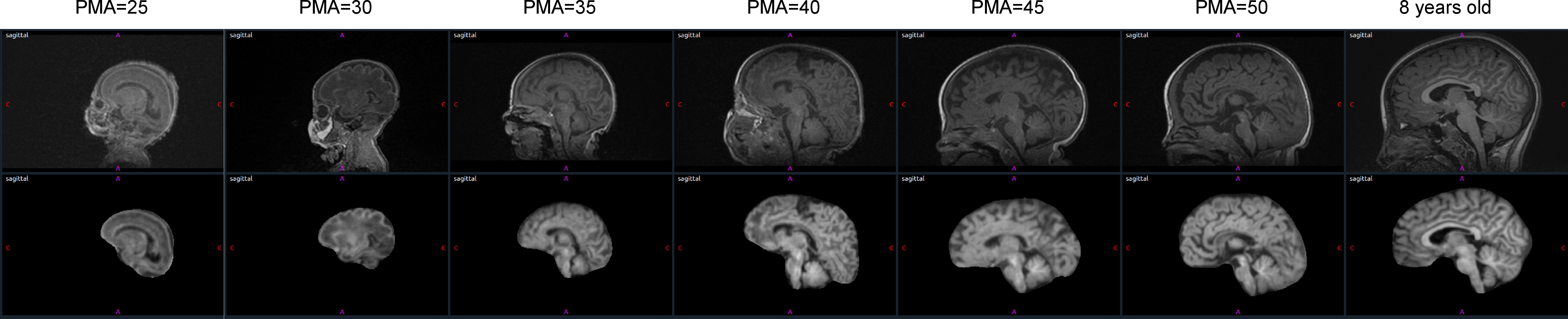}
\caption{Example of brain extraction and preprocessing of T1w MRI images using MGA-Net across various postmenstrual ages, ranging from neonatal (25 weeks to 50 weeks) to eight years old patients.}
\label{fig:exampleages}
\end{figure}

We rigorously validated MGA-Net using diverse MRI datasets, including those collected under less standardized conditions, demonstrating its robustness across varying data qualities (Figure \ref{fig:exampleages}). Furthermore, we used a dataset of real 3D ultrasound images from preterm neonates to validate MGA-Net’s preprocessing capabilities for ultrasound. This comprehensive approach not only enhances the utility of available data but also maximizes the effectiveness of preprocessing tasks across different imaging modalities.

\subsection{Related study}
\subsubsection{MRI}
In the context of traditional brain extraction tools, the Brain Extraction Tool (BET), introduced by Smith et al. \citep{smith2002fast}, remains a widely used benchmark for brain extraction. Although BET offers a robust, fast algorithm, it relies on a non-learning-based approach and often struggles with pathological cases and non-standardized imaging conditions, limiting its adaptability to various datasets. In contrast, recent advancements in deep learning have allowed for more flexible and generalizable methods.

Pei et al. \citep{pei2022general} propose an ensemble neural network (EnNet) based on 3D Convolutional Neural Networks (3DCNN) for skull stripping in multiparametric MRI scans. While their method demonstrates good performance across multiple imaging modalities, it is largely dependent on a comprehensive preprocessing, like noise reduction, bias correction and co-registration. In comparison, our approach does not rely on any of the preprocessing mentionend. Moreover, it can improve image quality by reconstructing the brain image.

Ranjbar et al. \citep{ranjbar2022weakly} address skull stripping in MRI images of brain tumor patients using the Dense-Vnet model. Their focus is limited to the specific pathological cases.
Yu et al. \citep{yu2022generalizable} introduce Brain Extraction Net (BEN), a domain-adaptive and semi-supervised deep neural network designed for cross-species and cross-modality brain extraction. BEN's domain-adaptability is one of its strengths; however, it requires a significant amount of domain-specific fine-tuning, which can be challenging in practical clinical settings. 
Hoopes et al. \citep{hoopes2022synthstrip} present SynthStrip, a learning-based tool that uses synthetic training data to handle diverse imaging protocols. SynthStrip eliminates the need for specific target contrasts during training, by using extensive synthetic data generation, which allows to generalize well to a variety of MRI datasets. Nevertheless, its reliance on synthetic data may introduce potential biases when applied to real clinical datasets, especially when dealing with neonatal brains.
Moreover, studies by Zhao et al. \citep{zhao2010automatic}, Gitler et al. \citep{gitler2017neurodegenerative}, and Radue et al. \citep{radue2015correlation} extend beyond brain extraction to neurodegenerative disease analysis, cortical surface reconstruction, and radiation therapy planning. These studies emphasize the critical role of accurate brain extraction in subsequent analyses. However, their primary focus on adult neuroimaging limits their applicability in neonatal and preterm infant populations, where brain structures are significantly different. Our MGA-Net, in contrast, is specifically designed to handle neonatal brain images, addressing these population-specific challenges while maintaining its genearl applicability accross adult population, as we have shown it in the subsequent sections.

\subsubsection{Ultrasound imaging}
In the field of ultrasound imaging, most recent studies have predominantly focused on fetal ultrasound, while neonatal imaging has received relatively less attention. Namburete et al. \citep{namburete2018fully} use a multitask learning model to extract the brain from fetal ultrasound images by employing a stack of 16 2D axial slices. Similarly, Moser et al. \citep{moser2020automated} adopt a 3D U-Net architecture to extract fetal brains from 3D ultrasound images in healthy fetuses between 14 to 31 weeks gestational age.

However, our work diverges from the fetal focus and concentrates on neonatal brain imaging, particularly for preterm infants. While Moser et al. \citep{moser2022bean} extend their work on fetal ultrasound to introduce the Brain Extraction and Alignment Network (BEAN) for 3D fetal brain ultrasound, we focus on developing automated brain extraction and preprocessing methods for neonatal ultrasound and MRI, a domain with limited existing literature. Addressing the gap in neonatal neuroimaging is critical to advancing the utility of ultrasound in this field.

\section{The Proposed MGA-Net}
We introduce MGA-Net, a novel deep learning architecture inspired by the U-Net framework for brain extraction and image preprocessing. MGA-Net is designed to handle both MRI and 3D ultrasound (US) images, incorporating modality-specific positional encoding and a mask-guided attention mechanism to enhance performance across these tasks.
\subsection{Network Architecture Overview}
MGA-Net features a U-Net backbone with two parallel decoders for brain extraction and image reconstruction. The input to the network can be either MRI or 3D US images. Inspired by Vaswani et al. \citep{vaswani2017attention}, sinusoidal positional encoding is integrated into the architecture to address the distinct spatial characteristics of each modality. It assigns specific values to MRI and US images (+1 for MRI and -1 for US) to help the network differentiate and process the modalities more effectively.
\subsection{Encoder-Decoder Design}
\begin{figure}[htpb]
\centering
\includegraphics[width=1\linewidth] {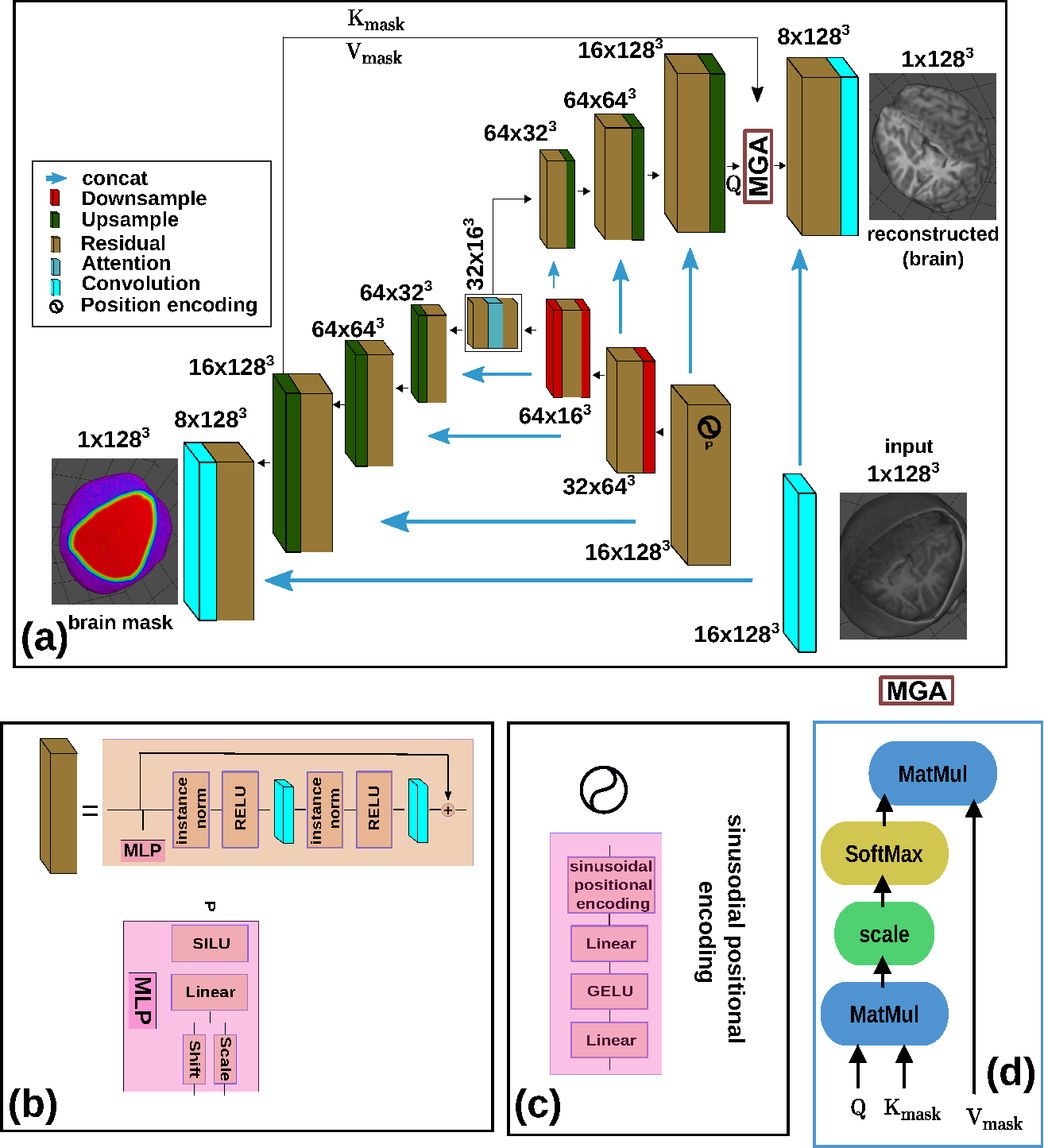}
\caption{(a) Architecture of the proposed preprocessing method for MRI and/or 3D Ultrasound images.
(b) Detailed architecture of the convolutional layers and MLP, highlighted in brown in part (a).
(c) Sinusoidal positional encoding mechanism.
(d) MGA layer: the Q matrix is obtained from the reconstruction branch, while the K and V matrices are derived from the mask branch.}
 \label{fig:mganet}
\end{figure}

The network encoder employs a series of convolutional layers for down-sampling, while up-sampling is performed using the nearest neighbor interpolation. The bottleneck layer incorporates an attention module, which enhances the model's ability to extract relevant features and improve robustness during preprocessing tasks.
MGA-Net features two distinct decoders: Brain Extraction Decoder and Image reconstruction decoder.
The first one is used to generate a probabilistic brain mask and the second one is used to reconstruct the image which has undergone noise removal, bias field correction, and non-brain region removal. The two decoders communicate to improve the model performance in both tasks.
Both outputs benefit from the integration of mask-guided attention, which ensures that the network captures the most relevant features during processing.Figure \ref{fig:mganet}
illustrates the full architecture of the proposed MGA-Net, highlighting the dual-decoder setup and attention mechanisms.

\subsection{Mask-Guided Attention Mechanism}
One of the key innovations of MGA-Net is the integration of a mask-guided attention module in the image preprocessing decoder. This module dynamically focuses on important image regions by assigning varying levels of importance to different parts of the image. The mask-guided attention consists of query (Q), key (K), and value (V) matrices:
Query (Q) is derived from the image reconstruction branch, incorporating global image context to guide attention.
Key (K) and Value (V) are derived from features in the brain extraction decoder, encoding spatial and structural information relevant to brain boundaries.
By combining local and global feature representations, mask-guided attention enhances the network's ability to focus on relevant regions, thereby improving brain extraction accuracy and overall image quality.
\subsection{Attention Mechanism Details}
In our implementation, we utilize a four-head attention mechanism to capture complex feature relationships and improve the network’s capacity to process diverse data types. This multi-head approach allows MGA-Net to capture a wide range of spatial and contextual features effectively.
\subsection{loss function}

The loss function $\mathcal{L}$ used in MGA-Net is a combination of tow main components, each addressing a specific aspect of the network's tasks. To ensure accurate boundary identification, we employ the mean squared error (MSE) loss denoted as $\mathcal{L}_{\text{boundary}}$, between generated boundary and the ground truth boundary. For the image reconstruction task, the $\mathcal{L}_{\text{recon}}$ is the sum of two loss function: 
the MSE loss $\mathcal{L}_{\text{MSE}}$ to measure pixel-wise accuracy between the reconstructed image and the reference brain-extracted image and the structural similarity index (SSIM) loss $\mathcal{L}_{\text{SSIM}}$ \citep{wang2004image}, which enhances the perceptual quality of the reconstructions by evaluating structural similarity between the reconstructed image and the reference brain-extracted image. Together, these loss components ensure both accurate segmentation and high-quality image reconstruction, and the overall loss function is the sum of the individual losses:
\begin{equation}
\mathcal{L} = \mathcal{L}_{\text{MASK}} + \mathcal{L}_{\text{MSE}} + \mathcal{L}_{\text{SSIM}}
\end{equation}

\subsection{Data Preprocessing}
The training mask output of the network has been constructed using a signed distance transform (SDT) method. It generates positive and negative contours from binary mask, similar to \citep{hoopes2022synthstrip}. The SDT assigns positive or negative values based on voxel proximity to the brain boundary, effectively defining non-strict brain regions.
\\
For both MRI and 3D US images, our preprocessing pipeline aims to extract brain, reduce noise, correct for artifacts, and enhance image quality, ensuring optimal ground truth for the network in training phase. The pre-processed images have been used as the reference brain-extracted image. A threshold of 4 mm is used in creating the reference brain extraction to build a brain mask with an additional boundary, which accommodates slight variations and inaccuracies in the brain's outer edges. This approach ensures a more robust and comprehensive representation of the brain structure, enhancing the model's performance and accuracy in identifying and segmenting brain regions.
MRI images undergo noise reduction using the Advanced Normalization Tools (ANTs) package with a shrink factor of 1, search radius of 2 and Rician noise model \cite{manjon2010adaptive}. 
N4 bias field correction is applied with a shrink factor of 2, 50 iterations set for three levels, a bias field full-width-at-half-maximum (FWHM) of 0.15, Wiener filter of 0.01, a maximum number of histogram bins of 200, a convergence threshold of 1e-3, B-spline grid spacing of 100 and spline order of 3.
Contrast-limited adaptive histogram equalization (CLAHE) is performed using scikit-image with a clip limit of 2 and a tile grid size of 8, enhancing image contrast and visibility of brain structures. 
For 3D ultrasound images, the preprocessing steps are similar to those used for MRI, except for the omission of bias field correction.
To standardize image dimensions, images with any dimension exceeding 128 voxels were resized so that the maximum dimension equaled 128, followed by zero-padding to achieve a final size of 128$\times$128$\times$128. For images initially smaller than 128 voxels in any dimension, zero-padding was applied directly to reach the 128$\times$128$\times$128 size. 
\subsection{Data Augmentation Strategy}
we employ an extensive data augmentation strategy tailored for both MRI and ultrasound images. 
One key technique is histogram matching, which adjusts the intensity distributions of input images to align with reference histograms, enhancing the model's ability to accurately segment non-brain regions. 
For T1-weighted and T2-weighted images, this process is exclusively applied based on pre-computed average histograms from the training sets, ensuring that reconstructed non-brain regions closely match ground truth segmentation masks. 
Additionally, we implement several other augmentation techniques: images are randomly rotated in the sagittal, axial, and coronal planes by angles ranging from -10° to 10° to simulate different head orientations, and zooming is performed by varying the image spacing from 0.5 to 4 while ensuring that the size remains within 50\% of the original dimensions. 
We also crop ground truth masks by varying the signed distance transform (SDT) threshold between 0 and 4, generating varying levels of brain boundary segmentation. 
To increase robustness against motion artifacts, we apply motion blur with kernel sizes of 5 and 11. 
Moreover, noise retained from preprocessing steps is added back to the images during training, simulating realistic noise patterns with a normal distribution to account for imaging artifacts. 
These augmentation techniques generate additional training samples and enhance the model’s ability to handle variability in real-world data, improving its performance in diverse imaging conditions.
\subsection{Final Image Reconstruction and Flexibility}
After preprocessing, MGA-Net processes images slightly larger than the brain boundaries, providing greater flexibility during brain extraction. Using an SDT threshold of 3, we ensure that the final reconstructed image retains accurate brain boundaries while also accommodating small deviations. This approach is particularly useful for handling complex, non-uniform brain regions that deviate from standard anatomical structures.
The final outputs from MGA-Net exhibit improved accuracy and robustness across both MRI and 3D US modalities, supporting more reliable preprocessing for downstream neuroimaging analyses.

\section{Experimental settings}

Our training dataset comprised a diverse assortment of medical images, totaling 710 distinct MRI sequences and 148 ultrasound images, as detailed in Table \ref{tab:combined_data}.
Given the limited availability of data from neonates, which is insufficient for training deep learning models, we incorporated various MRI sequences from multiple datasets. This strategy allows the proposed model to learn from a diverse range of data, thereby enhancing its generalizability.
The IXI dataset \footnote{Available at: http://brain-development.org/ixi-dataset} provides a comprehensive range of MRI contrasts and modalities, including T1-weighted (T1w), T2-weighted (T2w), proton density-weighted (PDw), magnetic resonance angiography (MRA), and diffusion-weighted imaging (DWI). We integrated this dataset into our training data using a randomly selected DWI image for training, akin to the approach detailed in Hoopes et al. \citep{hoopes2022synthstrip}. Additionally, data from the FSM dataset, as documented in Greve et al.'s work \citep{greve2021deep}, which includes standard acquisitions and quantitative T1 maps (qT1), were incorporated.
Furthermore, our training dataset includes pseudo-continuous arterial spin labeling (PCASL) scans acquired as stacks of 2D-EPI slices with low resolution and a limited field of view (FOV), a methodology elucidated in \citep{dai2008continuous,hoopes2022synthstrip}. We also included data from the QIN dataset drawn from previous studies \citep{mamonov2016data,clark2013cancer,prah2015repeatability}, consisting of pre-contrast clinical image stacks with thick slices from patients newly diagnosed with glioblastoma.
Finally, we have incorporated datasets from the University Hospital Puerta del Mar (HUPM), which include T1-weighted (T1w) images of preterm infants and 3D ultrasound (US) images. 
The ground-truth labels for HUPM dataset used in our study are derived through precise manual segmentation by experienced medical professionals. These labels represent brain boundaries and are critical for training and validating the deep learning model. Using specialized tools such as the MELAGE toolbox \citep{jafrasteh2023melage}, experts annotate key brain structures across multiple imaging planes (sagittal, coronal, and axial). This manual approach ensures high accuracy, especially in complex regions.

3D ultrasound images may not fully capture the entire brain region because of the narrow field of view of the anterior fontanel. Consequently, the ground truth mask may not represent total brain volume (TBV). We observed that between 5\% and 10\% of the brain volume can be missed by 3D US. To mitigate this issue, medical experts were recruited to segment the occluded parts to refine the ground truth masks. These occluded sections were segmented from three perspectives: sagittal, coronal, and axial views. Furthermore, to validate the results, the segmentation results were compared between two experts.

We extensively validated the proposed method using neonate datasets. Specifically, our validation dataset integrates 228 T1-weighted and 266 T2-weighted images from the Developing Human Connectome Project (DHCP), all obtained from neonatal subjects \citep{makropoulos2018developing}. Additionally, T1w and T2w images from the Albert neonate dataset \citep{gousias2012magnetic, gousias2013magnetic}, sourced from the repository maintained by Gousias et al. \footnote{Available at: https://brain-development.org/brain-atlases/neonatal-brain-atlases/neonatal-brain-atlas-gousias/}, were included in our validation efforts. We further enhanced our validation dataset with 70 neonate T1w images and 10 3D neonate ultrasound images from the University Hospital Puerta del Mar (HUPM), which were sourced from patients not included in the training dataset to ensure the robustness of our validation process. Finally, the BrainWeb dataset \citep{cocosco1997brainweb, collins1998design}, which includes 21 simulated T1 MRI images with the corresponding ground truth data, was used to validate the model's performance on adult MRI images.

We also included 712 ground truth TBV values from the HUPM dataset \citep{jafrasteh2023deep,benavente2021ultrasonographic} to evaluate the performance of the proposed model on 3D ultrasound images. The PMA of patients in these datasets ranged from 25 to 39 gestational weeks. Additionally, we conducted a comparative analysis in which MRI and US images obtained on the same day were processed using MGA-Net to estimate total brain volume. In total, there are 49 paired MRI and US images available for this analysis. The PMA of patients who underwent paired MRI and US was 25.4 to 52.7 weeks.
We evaluated the performance of the proposed method against BET \citep{smith2002fast}, Neural Preprocessing (NPP) \citep{he2023neural} and SynthStrip \citep{hoopes2022synthstrip} using the aforementioned datasets. Table \ref{tab:combined_data} summarizes the training and testing datasets.

\begin{table}[h]
\centering
\caption{Dataset, Modality, Resolution (mm), and Number of Samples for Training and Test Data}
\label{tab:combined_data}
\footnotesize
\setlength\tabcolsep{2pt}
\begin{tabular}{@{}lcccc@{}}
\toprule
\textbf{Dataset} & \textbf{Modality} & \textbf{Resolution (mm)} & \textbf{No. (Training)} & \textbf{No. (Test)} \\
\midrule
IXI  & T1w-T2w-PDw & 0.9 $\times$ 0.9 $\times$ 1.2 & 148 & - \\
 & MRA & 0.5 $\times$ 0.5 $\times$ 0.8 & 50 & - \\
 & DWI & 1.8 $\times$ 1.8 $\times$ 2.0 & 32 & - \\
FSM & T1w-T2w-PDw-qT1 & 1.0 $\times$ 1.0 $\times$ 1.0 & 138 & - \\
ASL & T1w MPRAGE & 1.0 $\times$ 1.0 $\times$ 1.0 & 43 & - \\
 & PCASL 2D-EPI & 3.4 $\times$ 3.4 $\times$ 5.0 & 43 & - \\
QIN & T1w-T2-FLAIR & 0.4 $\times$ 0.4 $\times$ 0.6 & 71 & - \\
 & T2w 2D-FSE & 1.0 $\times$ 1.0 $\times$ 5.0 & 39 & - \\
Infant  & T1w MPRAGE & 1.0 $\times$ 1.0 $\times$ 1.0 & 16 & - \\
HUPM & T1w - neonate & 0.9 $\times$ 0.9 $\times$ 0.9 & 78 & 70 \\
 & 3D US & 0.7 $\times$ 0.7 $\times$ 0.7 & 148 & 10 \\
 & T1w -8 years & 0.9 $\times$ 0.9 $\times$ 0.9 & 42 & - \\
BrainWeb & T1w & 1.0 $\times$ 1.0 $\times$ 1.0 & - & 21 \\
DHCP & T1w MRI & 0.5 $\times$ 0.5 $\times$ 0.5 & - & 228 \\
 & T2w MRI & 0.5 $\times$ 0.5 $\times$ 0.5 & - & 266 \\
Albert & T1w & 0.8 $\times$ 0.8 $\times$ 0.8 & - & 20 \\
 & T2w & 0.8 $\times$ 0.8 $\times$ 0.8 & - & 20 \\
\bottomrule
\end{tabular}
\end{table}

Supplementary Table \ref{tab:cnn_comp} presents the parameters of the model in details.
The number of batch sizes was set to $4$, and the network was trained over 1000 steps. 
All experiments were executed on a workstation equipped with two A5000 Nvidia GPUs, two Intel Xeon Gold 5220R CPUs, and 128 GB RAM.
The DICE coefficient, mean surface distance (MSD), recall, and accuracy metrics were used to assess segmentation accuracy. Furthermore, to validate the image reconstruction performance, the structural similarity index (SSIM) and peak signal-to-noise ratio (PSNR) were employed. Additionally, to evaluate total brain volume (TBV) estimation, the root mean square error (RMSE) and R-squared criteria were used.
As part of the ablation study, we compare the performance of the proposed method against several variants of the U-Net architecture: the standard U-Net, U-Net with data augmentation (U-Net DA), U-Net with sinusoidal positional encoding (U-Net SPE), and U-Net with mask-guided attention (U-Net MGA).

We also proposed a sensitivity analysis to assess the impact of varying the threshold parameter on the segmentation performance of the proposed method on different datasets.

\section{Results}

We compared the proposed methods BET, NPP, and SynthStrip on the test datasets considered in this study, as shown in Tables \ref{tab:comparison_dice_msd} and \ref{tab:comparison_recal_acc}. The results highlight the general superiority of the proposed method. Notably, our method achieved higher DICE coefficient (F1) scores and lower mean surface distance (MSD) values compared to the other methods, indicating better segmentation accuracy and spatial agreement with ground truth labels.
We compared the proposed method with BET, NPP, and SynthStrip across various test datasets, as detailed in Tables \ref{tab:comparison_dice_msd} and \ref{tab:comparison_recal_acc}, and evaluated the results using the DICE, MSD, Recall, and Accuracy metrics. The results consistently demonstrate the superior performance of the proposed method in most cases.

In the DHCP (T1) and HUPM (T1) datasets, the proposed method obtained significantly better results compared to the compared methods.
For the Albert (T1) and Albert (T2) datasets, the proposed method outperformed the others, demonstrating high segmentation accuracy and spatial agreement.
In the BrainWeb dataset, the proposed method was equal to SynthStrip in DICE but had a slightly higher MSD.

Segmentation accuracy and spatial agreement with ground truth labels are evaluated in Table \ref{tab:comparison_dice_msd} using the DICE coefficient (F1) and mean surface distance (MSD).
The proposed method achieved the highest DICE scores and lowest MSD values across most datasets, indicating superior performance:

The methods are compared in Table \ref{tab:comparison_recal_acc}  based on Recall and Accuracy.
In the DHCP (T1) dataset, the proposed method achieved high recall and accuracy, outperforming BET and NPP, and closely matching SynthStrip.
For the DHCP (T2) dataset, the proposed method matches NPP in recall but exhibits better accuracy.
In the HUPM (T1) dataset, our method excels in both Recall and Accuracy compared to the other methods.
For the Albert (T1) and Albert (T2) datasets, the proposed method again demonstrated high performance, surpassing BET and NPP and matching SynthStrip.
In the BrainWeb dataset, the proposed method had a slightly lower Recall compared to SynthStrip; however, it matched accuracy.

\begin{table}[H]
\centering
\caption{Performance comparison on various datasets using DICE and MSD criteria.}
\label{tab:comparison_dice_msd}
\scriptsize 
\setlength\tabcolsep{0.4pt} 
\begin{tabular}{lccccccccccc}
\toprule
\multirow{2}{*}{\textbf{Dataset}} & \multicolumn{2}{c}{\textbf{BET}} & \multicolumn{2}{c}{\textbf{Synthstrip}}  & \multicolumn{2}{c}{\textbf{NPP}} & \multicolumn{2}{c}{\textbf{Ours}} \\
\cmidrule(lr){2-3} \cmidrule(lr){4-5} \cmidrule(lr){6-7} \cmidrule(lr){8-9} \cmidrule(lr){10-11}
 & DICE(F1) & MSD & DICE(F1) & MSD & DICE(F1) & MSD & DICE(F1) & MSD\\
\midrule
DHCP (T1) & 0.82(0.07) & 8.82(3.02) & 0.94(0.02) & 3.88(0.83)  &0.86(0.03)& 8.22(1.35)& \textbf{0.96(0.01)}& \textbf{2.54(0.4)} \\
DHCP (T2) & 0.73(0.1) & 13.82(5.59) & \textbf{0.97(0.01)} & \textbf{1.84(0.34)} & 0.9(0.01) & 6.57(1.46) & 0.97(0.0)& 2.18(0.3)\\
HUPM (T1) & 0.75(0.09) & 7.44(2.32) & 0.89(0.06) & 2.9(1.22) &  0.69(0.09)  & 9.06(1.92) & \textbf{0.97(0.01)} & \textbf{0.94(0.19)}\\
Albert (T1) & 0.92(0.02) & 2.75(0.83) & 0.95(0.02) & 1.74(0.54) &  0.74(0.05) & 10.63(2.06)&  \textbf{0.98(0.0)} &  \textbf{0.74(0.13)} \\
Albert (T2) & 0.86(0.06) & 5.07(2.33) & 0.95(0.02) & 1.72(0.57) &  0.69(0.03) & 12.08(2.0) & \textbf{0.97(0.0)} & \textbf{1.01(0.15)} \\
BrainWeb & \textbf{0.97(0.01)} & \textbf{2.0(0.36)} & 0.97(0.01) & 2.05(0.37) & 0.93(0.01) &3.42(0.44)& 0.97(0.01) &  2.36(0.39) \\
\bottomrule
\end{tabular}
\end{table}

\begin{table}[H]
\centering
\scriptsize
\setlength\tabcolsep{0.4pt}
\caption{Performance comparison on various datasets using Recal and Accuracy criteria.}
\label{tab:comparison_recal_acc}
\begin{tabular}{lccccccccccc}
\toprule
\multirow{2}{*}{\textbf{Dataset}} & \multicolumn{2}{c}{\textbf{BET}} & \multicolumn{2}{c}{\textbf{Synthstrip}} & \multicolumn{2}{c}{\textbf{NPP}}& \multicolumn{2}{c}{\textbf{Ours}} \\
\cmidrule(lr){2-3} \cmidrule(lr){4-5} \cmidrule(lr){6-7}
\cmidrule(lr){8-9}
\cmidrule(lr){10-11}
 & \textbf{Recal} & \textbf{Accuracy} & \textbf{Recal} & \textbf{Accuracy} & \textbf{Recal} & \textbf{Accuracy}&\textbf{Recal} & \textbf{Accuracy}  \\
\midrule
DHCP (T1) & 0.71(0.1) & 0.88(0.04) & \textbf{0.99(0.01)} & 0.95(0.02)  &0.97(0.02)& 0.86(0.03)&  0.95(0.02) & 0.97(0.01)\\
DHCP (T2) & 0.59(0.13) & 0.82(0.06) & 0.95(0.02) & 0.97(0.01) & \textbf{0.97(0.01)}& 0.9(0.01)& \textbf{0.97(0.01)}& \textbf{0.98(0.0)}\\
HUPM (T1) & 0.99(0.01) & 0.74(0.11) & 0.99(0.01) & 0.91(0.05) & 0.99(0.01) & 0.66(0.12) & \textbf{0.98(0.02)} & \textbf{0.98(0.01)}\\
Albert (T1) & 0.97(0.02) & 0.93(0.02) & 0.99(0.01) & 0.96(0.02) &  1.0(0.0)&  0.71(0.08) & \textbf{0.98(0.01)} & \textbf{0.98(0.0)} \\
Albert (T2) & 0.83(0.1) & 0.89(0.04) & 0.99(0.01) & 0.96(0.02) &  0.99(0.01) &  0.62(0.05) & \textbf{0.99(0.01)} & \textbf{0.98(0.0)}\\
BrainWeb & 0.97(0.01) & 0.97(0.01) & \textbf{0.99(0.01)} & \textbf{0.97(0.0)}  & 1.0(0.0)&   0.94(0.01) &  0.98(0.01) &  0.97(0.0)\\
\bottomrule
\end{tabular}
\end{table}

For image reconstruction, the performance of the proposed method and the preprocessing pipeline mentioned in the previous section is compared in Table \ref{tab:psnr_ssim_comparison}, using PSNR and SSIM metrics across different datasets.

MGA-Net demonstrated better PSNR performance on the DHCP, Albert, and HUPM (US) datasets. For HUPM(T1) datasets, it showed lower PSNR but higher SSIM than the NPP method. NPP demonstrated better PSNR and SSIM values on brain web datasets.
Generally, NPP showed lower SSIM values than our proposed method.
It can be related to the training set in which it was trained on adult datasets rather than neonates.
The proposed method consistently achieved high PSNR and SSIM values compared to the pipeline approach, indicating superior image reconstruction quality.

In addition, Table \ref{tab:performance_metrics} provides performance evaluation using HUPM 3D ultrasound data, demonstrating high DICE (F1) score, low MSD, and high recall and accuracy rates.

\begin{figure}[h!]
\centering
\includegraphics[width=1\linewidth]{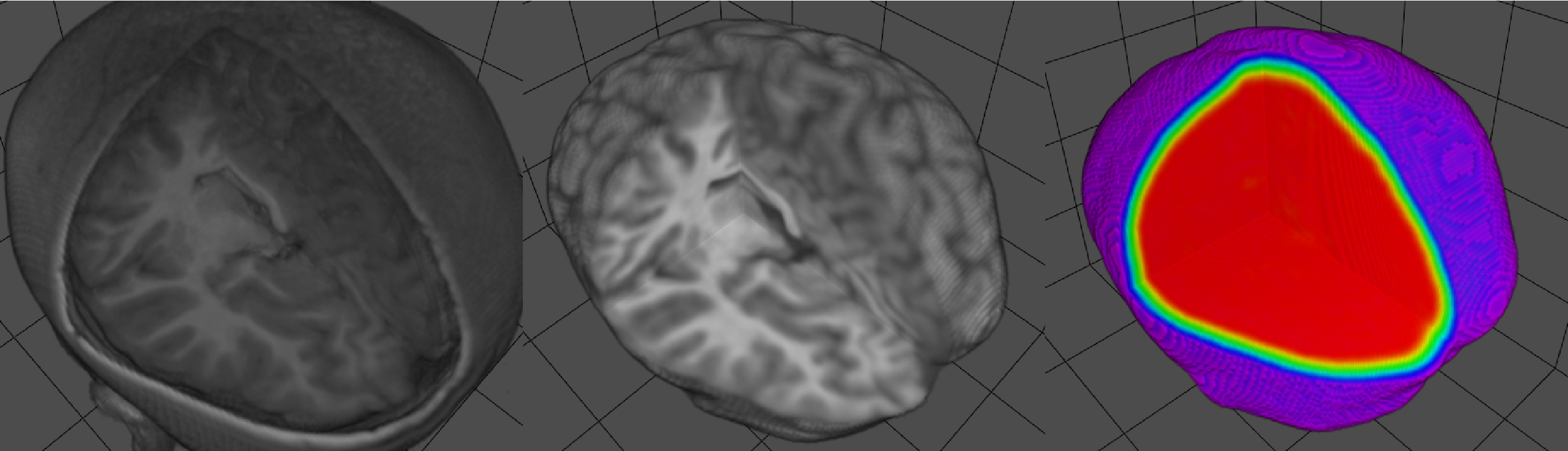}
\caption{From left to right: MRI image from an eight-year-old patient at HUPM, the reconstructed image using MGA-Net, and the brain boundary.}
 \label{fig:npmnetMRIEXAMPLE}
\end{figure}

\begin{figure}[h!]
\centering
\includegraphics[width=1\linewidth]{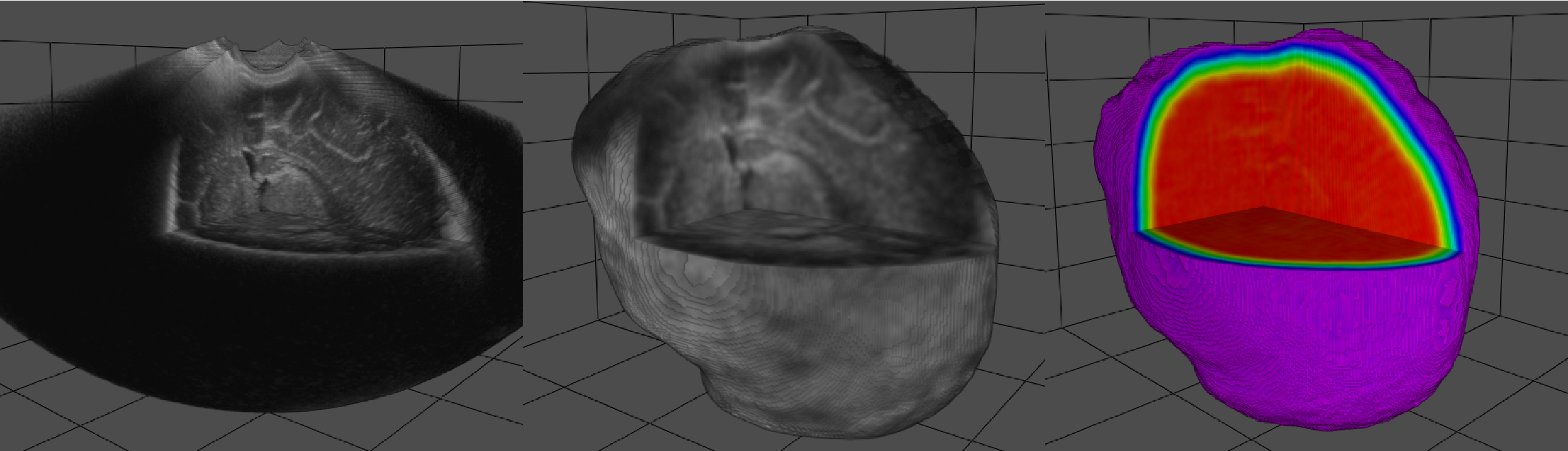}
\caption{From left to right: A 3D ultrasound image from a neonate at HUPM born at 32.4 gestational weeks and imaged at 32.7 weeks, the reconstructed image using MGA-Net, and the brain boundary.}
 \label{fig:npmnetUSEXAMPLE}
\end{figure}

Figure \ref{fig:npmnetUSEXAMPLE} illustrates a 3D ultrasound (US) image along with the corresponding reconstruction and segmentation outcomes obtained with the proposed MGA-Net and applied to a real ultrasound image sourced from the HUPM dataset. Visualization was performed using MELAGE software. Figure \ref{fig:npmnetMRIEXAMPLE} illustrates a 3D MRI image along with the corresponding reconstruction and segmentation outcomes obtained using the proposed MGA-Net.

Similarly, Figure \ref{fig:npmnetUS} illustrates the comparison between observed and predicted total brain volumes based on 3D ultrasound measurements at various postmenstrual ages (PMAs). The obtained RMSE and $R^2$ are 14.17 and 0.96, respectively, , indicating a strong correlation between observed and predicted total brain volume.

\begin{table}[H]
\centering
\scriptsize
\setlength\tabcolsep{0.4pt}
\caption{Comparison of performance between proposed method and preprocessing pipeline using PSNR and SSIM metrics across different datasets. PSNR and SSIM values are reported for both proposed method and pipeline approach. The values in parentheses represent the standard deviation.}
\label{tab:psnr_ssim_comparison}
\setlength\tabcolsep{3pt}
\begin{tabular}{lcccccccc}

\toprule
\multirow{2}{*}{\textbf{Dataset}} & \multicolumn{3}{c}{\textbf{PSNR}}  & \multicolumn{3}{c}{\textbf{SSIM}} \\
\cmidrule(lr){2-4} \cmidrule(lr){5-7}
 & \textbf{Pipeline} & \textbf{NPP} &
 \textbf{Ours} & \textbf{Pipeline} & \textbf{NPP}&
 \textbf{Ours} \\
\midrule
DHCP (T1)          & 18.56(1.02) & 25.06(1.87) & \textbf{30.26(2.32)} &0.76(0.05) & 0.91(0.02) & \textbf{0.97(0.01)}\\
DHCP (T2)          & 25.77(1.52) & 22.56(1.43)  & \textbf{30.33(1.2)} & 0.96(0.01) & 0.89(0.02) & \textbf{0.97(0.01)} \\
HUPM (T1)  & \textbf{34.54(3.39)}  & 33.95(1.7) & 32.88(5.21) & 0.99(0.0) & 0.88(0.05) & \textbf{0.99(0.01)} \\
Albert (T1)        & 19.51(2.95)  &28.36(1.85)&  \textbf{33.42(4.01)}&0.73(0.03) & 0.75(0.06) & \textbf{0.98(0.01)}   \\
Albert (T2)        & 30.86(2.35) & 27.15(1.42)  & \textbf{36.68(0.91)} & 0.95(0.02)  & 0.63(0.05) &  \textbf{0.97(0.0)}\\
BrainWeb (T1)      & 26.49(3.11)  & \textbf{29.82(3.94)}  &  25.65(3.08) &  \textbf{0.98(0.0)} &  0.98(0.01) &  0.97(0.01) \\
HUPM (US)          & 36.58(1.19)   &  - & \textbf{42.71(2.56)} & 0.99(0.0)   & - &  0.99(0.0) \\
\bottomrule
\end{tabular}
\end{table}

\begin{figure}[h!]
\centering
\includegraphics[width=1\linewidth]{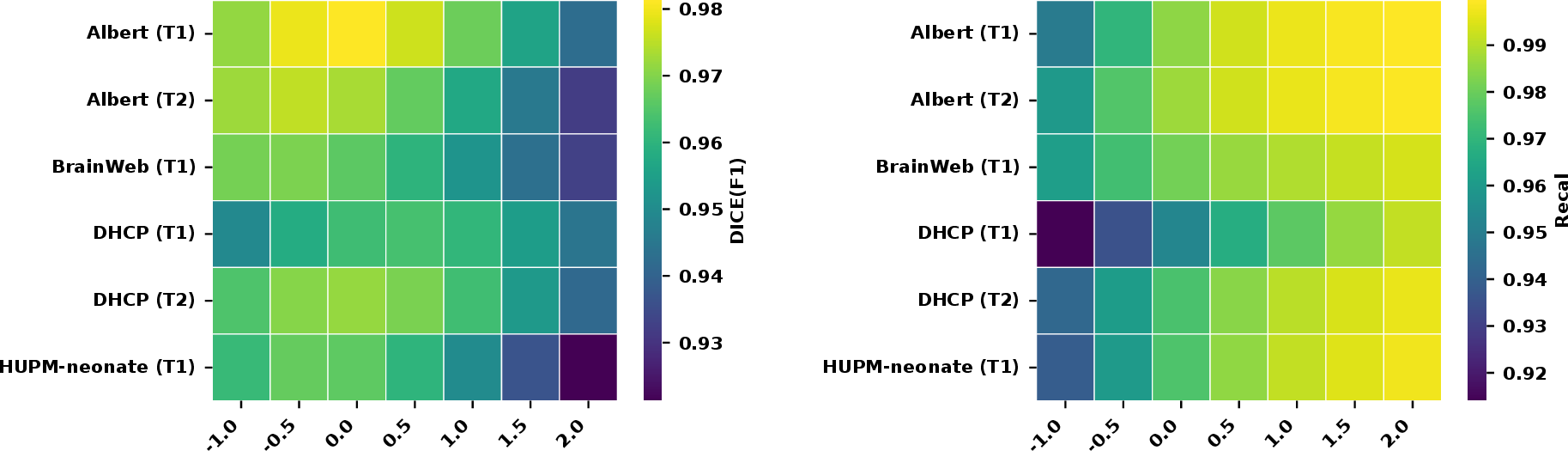}
\caption{Sensitivity analysis using DICE and Recal criteria for the proposed method with respect to changes in the threshold parameter across various datasets.}
 \label{fig:npmnet-sen}
\end{figure}

Figure \ref{fig:npmnet-sen} illustrates the results of the sensitivity analysis, where we systematically varied the threshold parameter and evaluated its effect on segmentation accuracy. By observing changes in performance metrics across different threshold values, we gain insights into how threshold parameters impact segmentation performance.
As threshold increases, the recal also increases, and the DICE coefficient decreases. Generally, a threshold equal to zero provides a balance between the dice coefficient and recall.
\\
As part of the ablation study, we compare the performance of the proposed method to that of the u-net architecture. These results demonstrate the higher segmentation accuracy of the proposed method.
Table \ref{tab:ablation} shows the comparison between U-Net 3D (ablated) and the proposed method on all test datasets in this study.

\section{Discussion}
The results consistently demonstrate the superior performance of the proposed method across most datasets and evaluation metrics (Table \ref{tab:method_comparison_all}). The significant improvement in DICE, MSD, Recall, and Accuracy metrics indicates that the proposed method outperforms BET, NPP, and SynthStrip in segmentation accuracy and spatial agreement with ground truth labels. This is particularly evident in the DHCP (T1), HUPM (T1), and Albert datasets, where our method consistently achieves higher scores.

The image reconstruction results highlight the robustness of MGA-Net in improving PSNR and SSIM metrics across various datasets. The proposed method demonstrated superior performance in the DHCP, Albert, and HUPM (US) datasets, while NPP performed better in the BrainWeb dataset. The lower SSIM values observed for NPP in neonates could be due to its training on adult datasets, suggesting that the proposed method is better suited for neonatal images.

\begin{figure}[h!]
\centering
\includegraphics[width=1\linewidth]{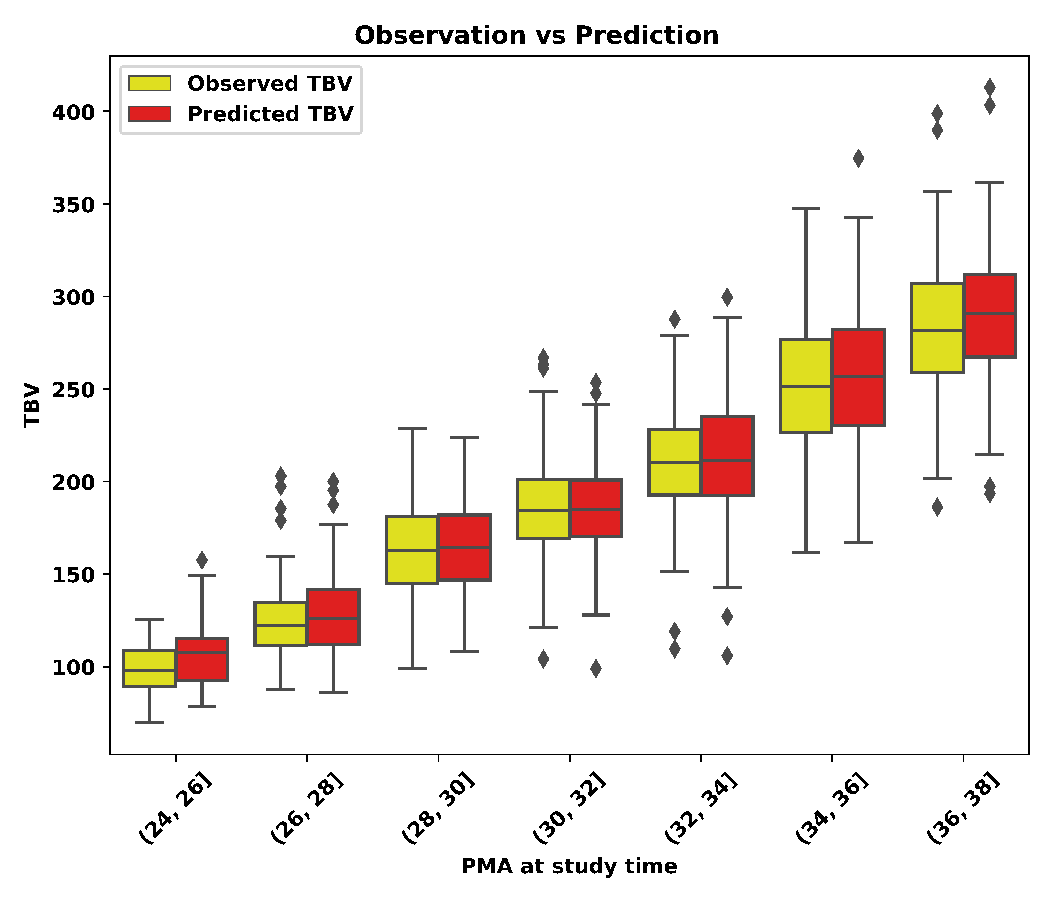}
\caption{Comparison between observed and predicted total brain volume based on 3D ultrasound measurements at various PMAs.}
 \label{fig:npmnetUS}
\end{figure}

The strong correlation between observed and predicted brain volumes, as shown in Figure \ref{fig:npmnetUS}, demonstrates the effectiveness of the proposed method in estimating total brain volume from 3D ultrasound images. This finding reinforces the potential of MGA-Net in clinical applications for neonatal brain analysis.
Figure \ref{fig:npmnetUS} demonstrates that the estimated total brain volume (TBV) values from 3D ultrasound (US) tend to overestimate in comparison to the corresponding MRI values, particularly at the lower range of TBV measurements.

Despite a few instances where other methods performed slightly better, the overall trend strongly supports the robustness and effectiveness of the proposed method in segmentation and reconstruction tasks. The architecture's ability to generalize across diverse datasets, including those with clinical abnormalities, further emphasizes its potential utility in real-world clinical settings.

\begin{table}[H]
\centering
\caption{Performance comparison on all the used datasets with different criteria.}
\label{tab:method_comparison_all}
\begin{tabular}{lcccc}
\toprule
\textbf{Method} & \textbf{BET} & \textbf{SynthStrip} & 
\textbf{NPP} &\textbf{Ours} \\
\midrule
DICE(F1) & 0.79(0.1) & 0.95(0.03) & 0.85(0.08)&  \textbf{0.97(0.01)} \\
MSD & 10.28(5.44)&  2.7(1.18)  &7.63(2.14)&  \textbf{2.11(0.65)} \\
Accuracy & 0.85(0.08)  &0.96(0.03)  &0.85(0.1)&  \textbf{0.98(0.01)} \\
Recal & 0.71(0.17) & 0.97(0.02) & 0.97(0.02)&  \textbf{0.97(0.02)} \\
\bottomrule
\end{tabular}
\end{table}

\begin{figure}[h!]
\centering
\includegraphics[width=1\linewidth]{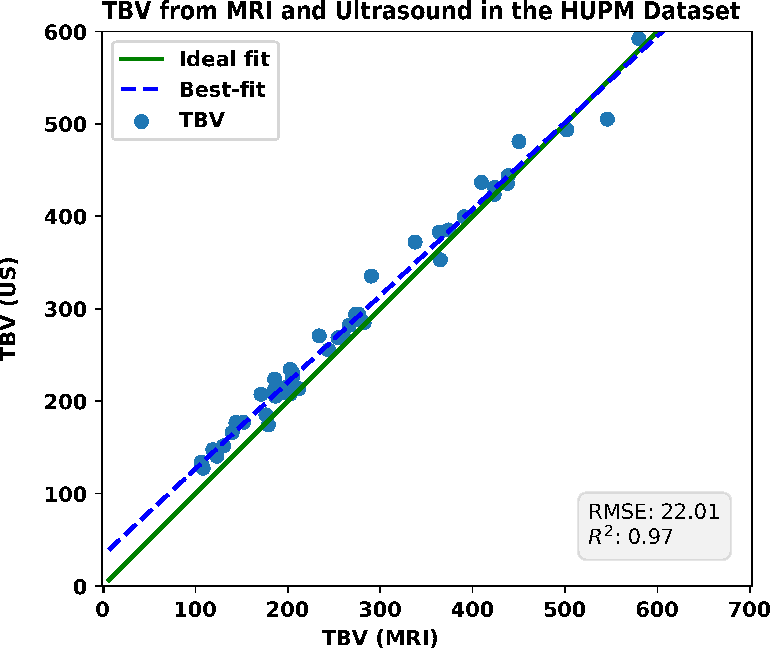}
\caption{Comparative Analysis of Estimated Total Brain Volume (TBV) from MRI and Ultrasound in the HUPM Dataset.}
 \label{fig:npmnetUS2}
\end{figure}

Figure \ref{fig:npmnetUS2} presents the results for 49 paired MRI and US images, where the RMSE is 22.01 and the $R^2$ value is 0.97, respectively. These results demonstrate a high degree of concordance between the TBV values obtained from these two distinct modalities.

\begin{table}[H]
\centering
\caption{Performance evaluation using HUPM 3D ultrasound data.}
\label{tab:performance_metrics}
\begin{tabular}{lcccc}
\toprule
\textbf{} & \textbf{DICE(F1)} & \textbf{MSD} & \textbf{Recal} & \textbf{Accuracy} \\
\midrule
\textbf{} & 0.95 (0.02) & 1.49 (0.45) & 0.96 (0.02) & 0.96 (0.01) \\
\bottomrule
\end{tabular}
\end{table}

\begin{table}[H]
\centering
\caption{Results of ablation study}
\label{tab:ablation}
\begin{tabular}{lcccc}
\toprule
\textbf{} & \textbf{DICE(F1)} & \textbf{MSD} & \textbf{Recal} & \textbf{Accuracy} \\
\midrule
\textbf{U-Net} &  0.94(0.02) & 3.40(1.27) &  0.90(0.02)  & 0.95(0.04)\\
\textbf{U-Net (DA)} &  0.94(0.02) & 3.18(1.17) &  0.90(0.04)  & 0.95(0.01)\\
\textbf{U-Net (SPE)} &  0.94(0.02) & 3.31(1.25) &  0.90(0.05)  & 0.95(0.02)\\
\textbf{U-Net (MGA)} &  0.94(0.02) & 3.02(1.17) &  0.90(0.04)  & 0.95(0.02)\\
\textbf{Ours} &  \textbf{0.97(0.01)} & \textbf{2.11(0.65)}  & \textbf{0.98(0.01)} & \textbf{0.97(0.02)} \\
\bottomrule
\end{tabular}
\end{table}

Additionally, We examined the impact of incorporating ultrasound (US) images on MRI segmentation performance. Our results show that adding US images does not significantly enhance the performance of the proposed method. This lack of improvement may be due to the limited number of US images in the training sets, which may not provide sufficient information to boost MRI segmentation accuracy. Despite this, the proposed algorithm is designed to effectively process both MRI and US images. While the current dataset may not fully leverage the potential benefits of integrating US images, the ability of the model to handle both image types suggests that with more extensive and diverse datasets, integrating US images could contribute to developing a more robust and versatile model. Further research with larger and more varied datasets is needed to fully understand the potential advantages of combining MRI and US images.

\subsection{image anomalies}
\begin{figure}[h!]
\centering
\includegraphics[width=1\linewidth]{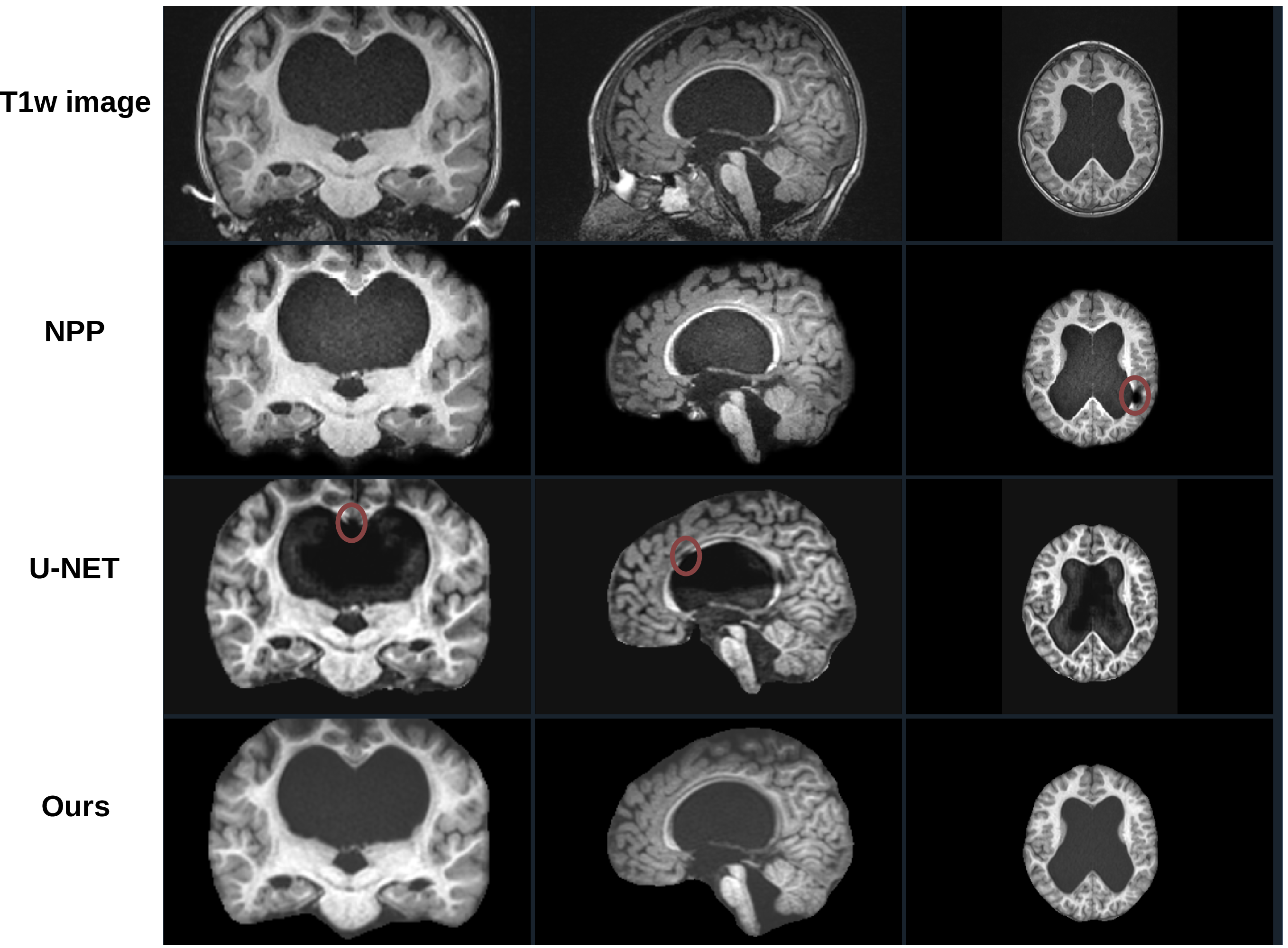}
\caption{Coronal, sagital and axial views (from left to right) of an eight-year-old preterm-born patient. The top image is the initial image.}
 \label{fig:ablation}
\end{figure}

Figure \ref{fig:ablation} presents the three views of an eight-year-old preterm-born patient with reconstruction using NPP, U-Net, and the proposed method. As can be seen, the proposed method can successfully reconstruct the image even though there is an anomaly in the image, whereas NPP and U-Net have some problems in reconstructing the entire image (red circle in Figure \ref{fig:ablation}.

\begin{figure}[h!]
\centering
\includegraphics[width=0.7\linewidth]{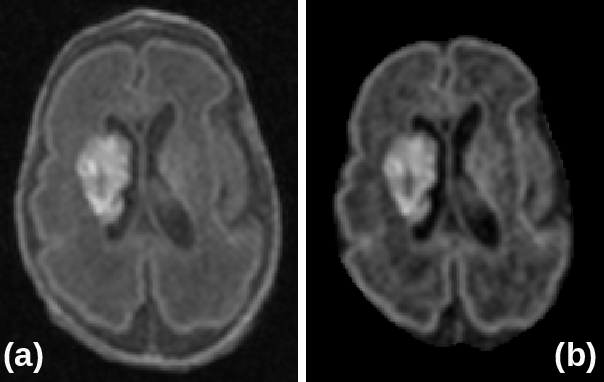}
\caption{a) Axial section of a 3D MRI from the HUPM dataset, showing the brain at 29 weeks of postmenstrual age (PMA). This image showcases typical neuroanatomical features and potential abnormalities often found in preterm neonates, providing a baseline for preprocessing assessment. b) The same image post-application of the MGA-Net, demonstrate the effectiveness of the network in enhancing image clarity and detail for better clinical assessment.}
 \label{fig:lesion}
\end{figure}

The dataset obtained from Hospital Universitario Puerta del Mar (HUPM) originates from clinical settings, involving preterm neonates who may present with neurological abnormalities, such as abnormal tissue structures or atypical brain morphology (Figure \ref{fig:lesion}). Some of these patients have been incorporated into the training dataset, while others are allocated to the test dataset to evaluate the model's generalizability and robustness.
Despite the inherent challenges associated with these variations, the results demonstrate that the performance of the proposed method on the test dataset is superior to that of conventional methods. This indicates that the proposed method can be effectively accommodated and adjusted for clinical abnormalities in its analyses. We attribute this robustness to the network architecture, which enhances its ability to generalize across uncommon and variant anatomical features.

\begin{figure}[h!]
\centering
\includegraphics[width=1\linewidth]{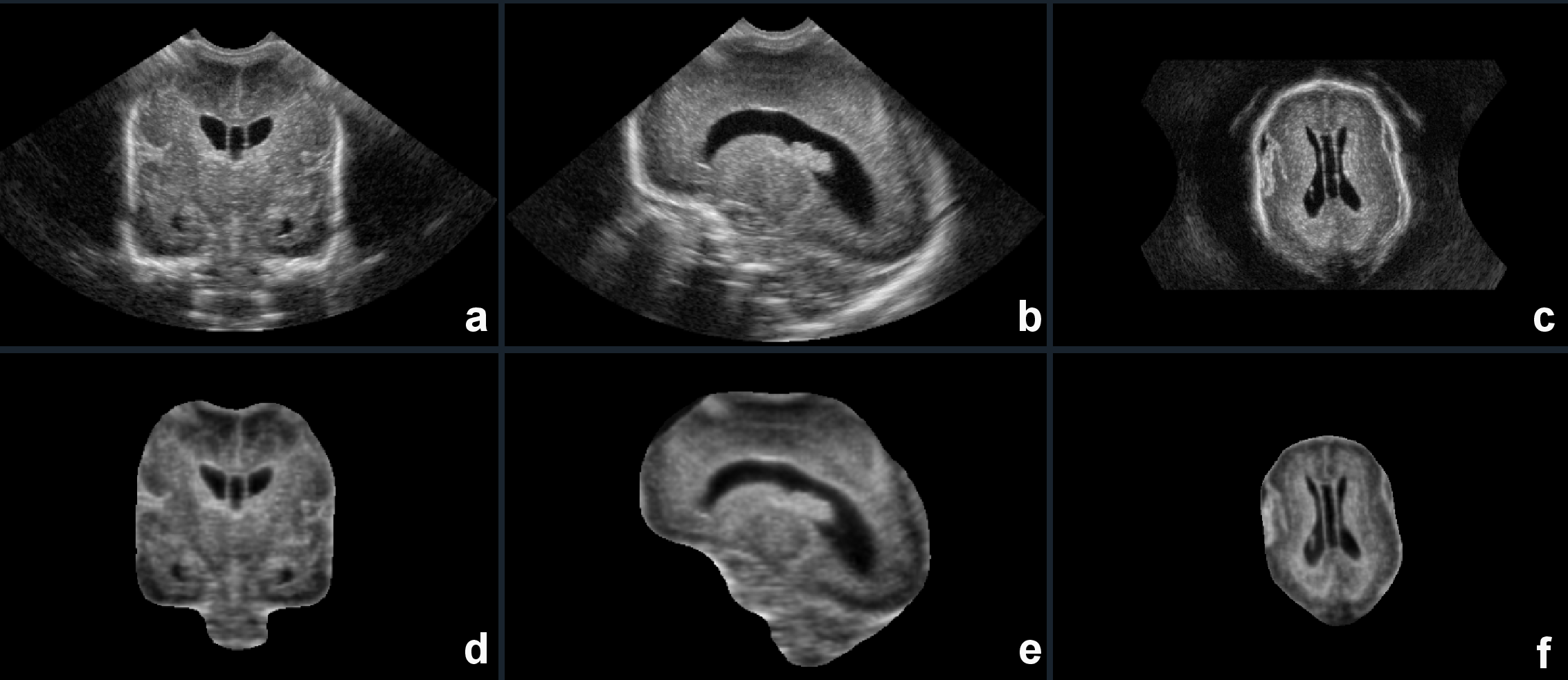}
\caption{3D Ultrasound (US) image of a preterm neonate born at 28.29 gestational weeks, imaged at 29 weeks: Top) Original US image, Bottom) US image processed with MGA-Net. Coronal, sagittal, and axial sections are displayed from left to right.}
 \label{fig:us_validation}
\end{figure}

For the ultrasound (US) datasets, Figure \ref{fig:us_validation} presents coronal, sagittal, and axial view of a 3D ultrasound from the HUPM dataset, displaying the image before and after processing with the MGA-Net. This figure illustrates the effectiveness of the proposed method in reconstructing images, highlighting how MGA-Net enhances the visual quality and detail for better diagnostic evaluation.

\section{Conclusions}

This study introduced the mask-guided attention neural network (MGA-Net), a novel framework designed to tackle the inherent complexities of preprocessing MRI and 3D ultrasound images. Our comparative analysis highlights the superiority of MGA-Net over other state-of-the-art methods, such as BET, NPP, and SynthStrip. The results demonstrate that MGA-Net not only enhances the precision of brain extraction and segmentation and excels at reconstructing high-quality images.
Significantly, MGA-Net demonstrated exceptional ability to estimate total brain volume from 3D ultrasound images, which is an area in which existing methods often falter. A comprehensive sensitivity analysis of the threshold parameters further demonstrates the robustness of the model, ensuring reliability even under varied operational conditions.
Although the inclusion of sinusoidal positional encoding, mask-guided attention, or data augmentation individually does not enhance model performance, our study demonstrates that combining all these components significantly improves the model’s performance. Sinusoidal positional encoding captures spatial relationships within images, while mask-guided attention allows the network to focus on relevant regions, improving preprocessing and segmentation. Additionally, extensive data augmentation, including a novel histogram-based approach, enhances the method's robustness. Collectively, these components enable MGA-Net to handle the diverse statistical properties of MRI and ultrasound images more effectively, leading to superior performance across multiple critical metrics.
Overall, MGA-Net is a freely available tool for enhancing the accuracy and efficiency of neuroimaging data analysis across diverse applications in research and clinical settings.

\section*{Code Availability}
The network code and weights are available through \href{https://github.com/BahramJafrasteh/MGA-Net}{https://github.com/BahramJafrasteh/MGA-Net }

\section*{Acknowledgements}

This study was funded by the Instituto de Salud Carlos III (ISCIII) through the project "DTS22/00142" and co-funded by the European Union.
We acknowledge the use of data from the developing Human Connectome Project (dHCP), KCL-Imperial-Oxford Consortium, funded by the European Research Council under the European Union Seventh Framework Program (FP/2007-2013) / ERC Grant Agreement no. [319456]. We thank the families who generously supported this trial. The DCHP data were used to evaluate model performance.
This study was reviewed and approved by the Local  Research Ethic Committee (registration number:139.22; approval date: December 19, 2022).
%
%
%
\clearpage
\bibliographystyle{elsarticle-num-names}
\bibliography{mybibliography}

\pagebreak
\appendix

\renewcommand{\thesection}{S\arabic{section}}
\renewcommand{\thefigure}{S\arabic{figure}}
\renewcommand{\thetable}{S\arabic{table}}
\counterwithin{figure}{section}
\counterwithin{table}{section}

\section{Supplementary Material}

\begin{table}[H]
  \centering 
  \caption{parameters of the proposed model. The decoder parameters of the brain extraction and image reconstruction are the same.}
  \resizebox{\textwidth}{!}{
  \begin{tabular}{cccc}
  \toprule
    \textbf{Layer} & \textbf{Specifications} & \textbf{Parameters} & \textbf{Output Dimension} \\
    \midrule
    \textbf{Input Image} & & & (1,128,128,128) \\
    \midrule
    \textbf{Encoder 1} & Kernel size 3, stride 1 & (16,1,3,3,3) & (16,128,128,128) \\ 
    \midrule
    \textbf{Encoder 2.1} & Kernel size 3, stride 1 & (16,16,3,3,3) & (16,128,128,128) \\ 
    \textbf{Encoder 2.2} & Kernel size 1, stride 1 & (16,16,1,1,1) & (16,128,128,128) \\ 
    \midrule
    \textbf{Encoder 3.1} & Kernel size 1, stride 2 & (16,16,1,1,1) & (16,64,64,64) \\ 
    \midrule
    \textbf{Encoder 4.1} & Kernel size 3, stride 1 & (32,16,3,3,3) & (32,64,64,64) \\ 
    \textbf{Encoder 4.2} & Kernel size 1, stride 1 & (32,32,1,1,1) & (32,64,64,64) \\ 
    \textbf{Encoder 4.3} & Kernel size 1, stride 1 & (32,16,1,1,1) & (32,64,64,64) \\ 
    \midrule
    \textbf{Encoder 5.1} & Kernel size 1, stride 2 & (32,32,1,1,1) & (32,32,32,32) \\ 
    \midrule
    \textbf{Encoder 6.1} & Kernel size 3, stride 1 & (64,32,3,3,3) & (64,32,32,32) \\ 
    \textbf{Encoder 6.2} & Kernel size 1, stride 1 & (64,64,1,1,1) & (64,32,32,32) \\ 
    \textbf{Encoder 6.3} & Kernel size 1, stride 1 & (64,32,1,1,1) & (64,32,32,32) \\ 
    \midrule
    \textbf{Encoder 7.1} & Kernel size 1, stride 2 & (64,64,1,1,1) & (64,16,16,16) \\ 
    \midrule
    \textbf{Bottleneck 8.1} & Kernel size 3, stride 1 & (64,64,3,3,3) & (64,16,16,16) \\ 
    \textbf{Bottleneck 8.2} & Kernel size 1, stride 1 & (64,64,1,1,1) & (64,16,16,16) \\ 
    \midrule
    \textbf{Attention} & head dimension 16, number of heads 4&&\\
    \textbf{Attention 2.1} & Kernel size 1, stride 1 & (192,64,1,1,1) & (192,16,16,16) \\ 
    \textbf{Attention 2.2} & Kernel size 1, stride 1 & (32,64,1,1,1) & (32,16,16,16) \\ 
    \midrule
    \textbf{Decoder 1.1} & Kernel size 3, stride 1 & (32,32,3,3,3) & (32,16,16,16) \\ 
    \textbf{Decoder 1.2} & Kernel size 1, stride 1 & (32,32,1,1,1) & (32,16,16,16) \\ 
    \midrule
    \textbf{Decoder 2.1} & Kernel size 3, stride 1 & (32,64,3,3,3) & (32,32,32,32) \\ 
    \textbf{Decoder 2.2} & Kernel size 1, stride 1 & (32,32,1,1,1) & (32,32,32,32) \\ 
    \textbf{Decoder 2.3} & Kernel size 1, stride 1 & (32,64,1,1,1) & (32,32,32,32) \\ 
    \midrule
    \textbf{Decoder 3.1} & Kernel size 3, stride 1 & (16,64,3,3,3) & (16,64,64,64) \\ 
    \textbf{Decoder 3.2} & Kernel size 1, stride 1 & (16,16,1,1,1) & (16,64,64,64) \\ 
    \textbf{Decoder 3.3} & Kernel size 1, stride 1 & (16,64,1,1,1) & (16,64,64,64) \\ 
    \midrule
    \textbf{MGA} & head dimension 16, number of heads 4& & \\ 
    \midrule
    \textbf{Decoder 4.1} & Kernel size 3, stride 1 & (8,32,3,3,3) & (8,128,128,128) \\ 
    \textbf{Decoder 4.2} & Kernel size 1, stride 1 & (8,8,1,1,1) & (8,128,128,128)\\ 
    \textbf{Decoder 4.3} & Kernel size 1, stride 1 & (8,32,1,1,1) & (8,128,128,128) \\ 
    \midrule
    \textbf{Last convolution} & Kernel size 3, stride 1 & (1,8,3,3,3) & (1,128,128,128) \\ 
    \bottomrule
  \end{tabular}
  \label{tab:cnn_comp} 
  }
\end{table}

\end{document}